\documentclass[10pt]{article}
\usepackage{amsmath}
\usepackage{amssymb}
\usepackage{graphicx}
\usepackage{cite}
\usepackage{color}

\usepackage{CJK}

\usepackage[labelfont=bf,labelsep=period,justification=raggedright]{caption}
\makeatletter

\renewcommand{\@biblabel}[1]{\quad#1.}
\makeatother
\date{}
\begin{document}
\begin{CJK}{UTF8}{gbsn}
\begin{flushleft}
{\footnotesize (Version 0.1)}\\[2ex]

 {\Large
  \textbf{The Optimal Size of Stochastic Hodgkin-Huxley Neuronal Systems for Maximal Energy Efficiency in Coding of Pulse Signals}
 }
\\[1ex]
 Lianchun Yu$^{1,2 \ast}$, Liwei Liu$^{3}$
\\[1ex]

\bf{1} Institute of Theoretical Physics, Lanzhou University, Lanzhou 730000, China
\\
\bf{2} Key Laboratory for Magnetism and Magnetic Materials of the
Ministry of Education, Lanzhou University, Lanzhou 730000, China
\\
\bf{3} College of Electrical Engineering, Northwest University for
Nationalities, Lanzhou 730070, China
\\[1ex]

$\ast$ E-mail: yulch@lzu.edu.cn
\end{flushleft}

\section*{Abstract}
The generation and conduction of action potentials represents a
fundamental means of communication in the nervous system, and is a
metabolically expensive process. In this paper, we investigate the
energy efficiency of neural systems in a process of transfer pulse
signals with action potentials. By computer simulation of a
stochastic version of Hodgkin-Huxley model with detailed description
of ion channel random gating, and analytically solve a bistable
neuron model that mimic the action potential generation with a
particle crossing the barrier of a double well, we find optimal
number of ion channels that maximize energy efficiency for a neuron.
We also investigate the energy efficiency of neuron population in
which input pulse signals are represented with synchronized spikes
and read out with a downstream coincidence detector neuron. We find
an optimal combination of the number of neurons in neuron population
and the number of ion channels in each neuron that maximize the
energy efficiency. The energy efficiency depends on the characters
of the input signals, e.g., the pulse strength and the inter-pulse
intervals. We argue that trade-off between reliability of signal
transmission and energy cost may influence the size of the neural
systems if energy use is constrained.

\section*{Author Summary}
Action potentials are the basic way that information is transmitted
in the brain. Those electrical 'spikes' also cost a large fraction
of metabolic energy in brains. It is believed the neural systems
should be evolved to be energy efficient under the pressure of
natural selection. The demand for energy efficiency could be large
enough to influence the design of the neural system. Since the
neural systems are basically stochastic devices, it is interesting
to investigate the effects of noise on the energy efficiency of
neural systems. In this paper, using computer simulation and
analytical methods, we showed that the energy efficiency exhibits
maxima corresponding to optimal number of ion channels in the neuron
and optimal number of neurons in a pulse signal detection task.
Those results indicate that the size of the neural systems may be
optimized to transfer information energy efficiently.
\section*{Introduction}
Neural processing is metabolically
expensive~\cite{Richter_Metabolism_2001}. The human brain accounts
for about 20\% of an adult's resting metabolic rate. A large
fraction of this energy is used for action potential generations,
which rely on the potential energy stored in transmembrane ion
gradients\cite{attwell_energy_2001}. Though action potential itself
does not consume energy, maintaining this ionic concentration
gradients and restoring them after an action potential requires
energy released during ATP hydrolysis to drive ATPase
$Na^{+}$/$K^{+}$
exchangers~\cite{Magistretti11092009,Siekevitz15102004}. These
metabolic demands could be large enough to influence the design,
function and evolution of brains~\cite{laughlin_metabolic_1998}.
Therefore, it is believed that under the pressure of natural
selection, the neurons, neural coding and neuronal circuit will be
evolved to be energy efficient if metabolic energy is
limited~\cite{Niven01062008,Laughlin:2001p42719}.

In the process of action potential generation, the inward $Na^+$
current depolarizes the membrane and generates the upstroke, whereas
the outward $K^+$ current repolarizes the membrane and facilitates
the downstroke. The overlap of those two ion fluxes would result in
an electrically neutral exchange of positive ions, wasting $Na^+$
and accordingly energy~\cite{Margineanu}. Thus it is suggested
minimizing the overlap of those two ion fluxes will improve the
energy efficiency for action potential
generation~\cite{10.1371/journal.pcbi.1002456}. Recent experimental
investigation on the nonmyelinated mossy fibers of the rat
hippocampus supported this suggestion~\cite{Alle11092009}.

The energy efficiency of the neural system is also dependent on the
size of the neuron~\cite{Sarpeshkar:1998:AVD:303222.303223}. For
example, one would expect that neurons with small size will be more
efficient because less ion channels are involved, thus less ion
exchange through ion pump. However, the ion channels are stochastic
devices, less number of ion channels in the neurons will lead to
larger fluctuations in the transmembrane ion
current~\cite{Steinmetz99subthresholdvoltage}. This so called ``ion
channel noise" has been demonstrated not only give variability in
the response of neuron to external stimuli, but also cause
spontaneous action potentials, and damage the reliability of signal
processing~\cite{Chow19963013, Laughlin:1989:RSN:103938.103955}. In
this case, trade-offs between information transfer and energy use
could strongly influence the number of ion channels used by the
neurons.

An elegant theoretical analysis, using a simple model for the
generation of graded electrical signals by sodium and potassium
channels, showed that there exists an optimum number of ion channels
that maximize energy efficiency. The optima depend on several
factors: the relative magnitudes of the signaling cost (current flow
through channels), the fixed cost of maintaining the system, the
reliability of the input, additional sources of noise, and the
relative costs of upstream and downstream
mechanisms\cite{Schreiber02energy-efficientcoding}. While this is
true, as discussed above, instead of graded electrical signal, it is
the action potentials that carry the information in the neural
systems, and contribute to major cost of energy. Therefore, it is
interesting to study the energy efficiency of neurons which code
signals digitally with action potentials when ion channel noise is
concerned.

  Another interesting question is if this trade-offs between information transfer and energy use could influence the size of the neural network.
It is well known that the information process in brain is
accomplished with a group of neurons working cooperatively, not with
only one neuron~\cite{bear_neuroscience:_2007}. Studies have shown
that neurons could be synchronized to prevent from noise
perturbation and facilitating
  reliability of information transmission\cite{10.1371/journal.pcbi.1000637, PhysRevE.78.051909}. However, large number of
  synchronized spikes generated in the neural systems will give a
  large energy burden on them. In this case, it is
  possible that the number of neurons in the neural system is
  optimized to balance the reliability of information transfer and
  energy cost.

In this paper, we employ analytical and computational methods to
calculate the metabolic energy efficiency of single neuron and
neuron population performing a pulse-signal detection task in the
environment of ion channel noise. A simplified model of a
one-dimensional bistable Langevin equation, which mimics the action
potential generations with a particle crossing the barrier of a
double well, are solved analytically for the pulse signal detection
rate and spontaneous firing rate. Based on those results, the energy
efficiency of single neuron are investigated. Results show that for
the single neuron, there is an optimal number of ion channels
corresponding to maximal energy efficiency in detection of pulse
signals. The optima depend on the signal properties, e.g., pulse
strength, inter-pulse interval. To investigate the energy efficiency
of neuron population, the pulse inputs are applied to a group of
neurons, the outputs of those neurons are received and read out by a
downstream coincidence detector (CD) neuron. In this case, the
reliability of detection is enhanced, and there is an optimal
combination of the number of neurons and the number of ion channels,
with which the energy efficiency is maximized. The optimal number of
neurons depends on the threshold of downstream coincidence detector
neuron. Those results are confirmed with computer simulation of a
stochastic Hodgkin-Huxley (HH) model in which the detailed
description of ion channel random gating are considered.

\section*{Results}
\subsection*{Threshold fluctuation and spontaneous firings }
\begin{figure}[h]
\begin{center}
\includegraphics[width=0.850\textwidth]{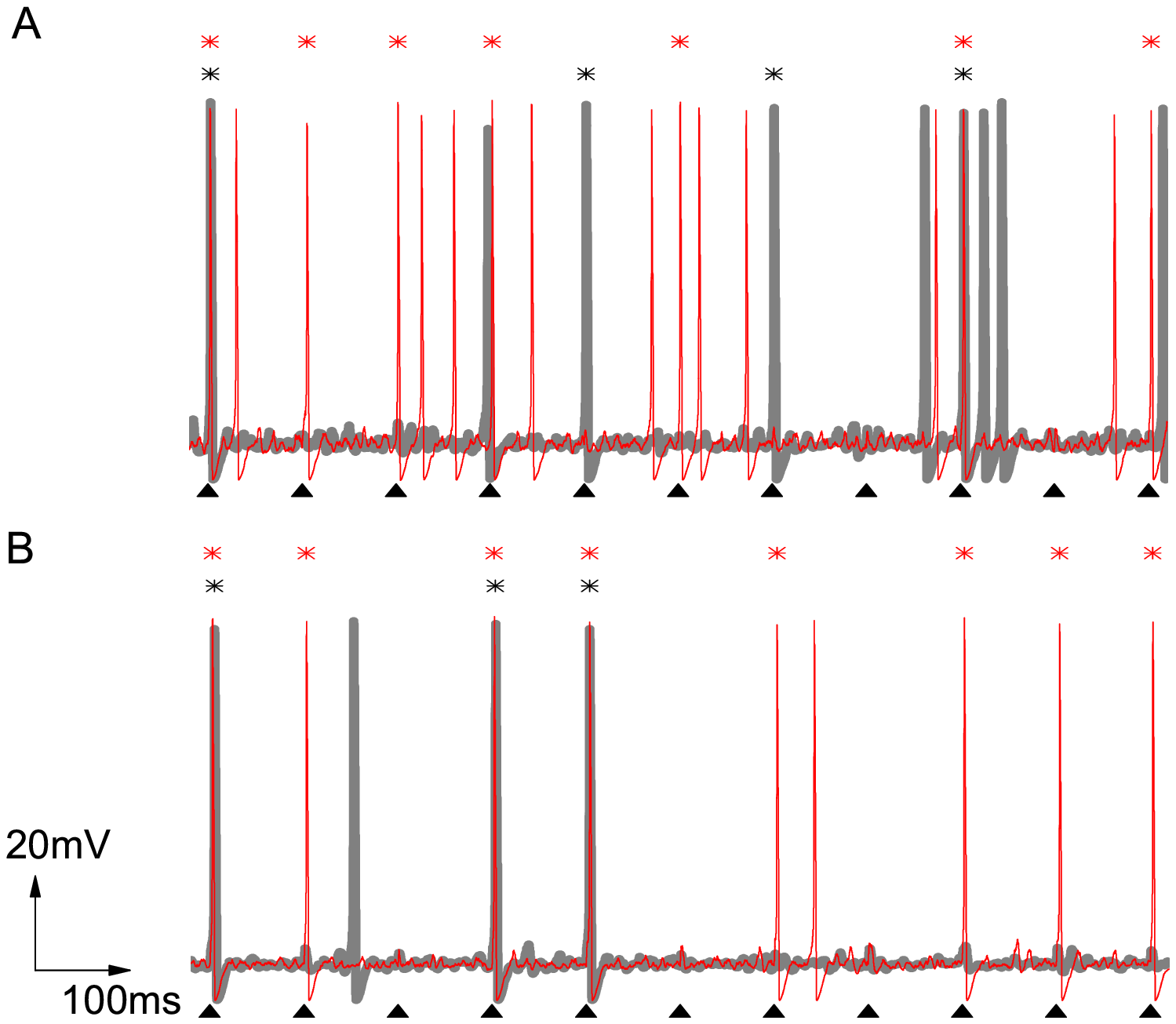}
\end{center}
\caption{ {\bf Demonstration of membrane potential traces for
stochastic HH models receiving pulse inputs.} The membrane area of
the neuron is $120 \mu m^2$ in (A) and $240 \mu m^2$ in (B). Wide
gray lines correspond to
 input pulses with strength $I=5~\mu A /cm^{2}$ and
red lines $I=8~\mu A /cm^{2}$. The black triangles mark the time the
pulse inputs applied. The black stars mark the APs in response to
pulse inputs with strength $I=5~\mu A /cm^{2}$ and red stars
$I=8~\mu A /cm^{2}$.} \label{Demo}
\end{figure}

\begin{figure}[h]
\begin{center}
\includegraphics[width=0.85\textwidth]{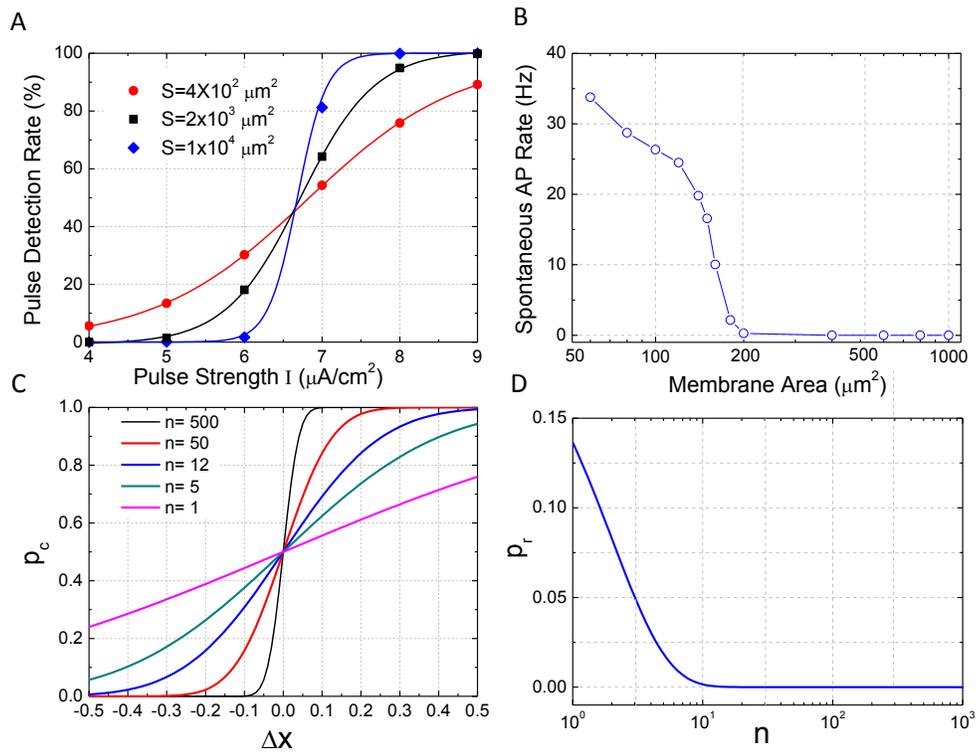}
\end{center}
\caption{ {\bf The pulse detection rate and spontaneous firing rate
of neurons with ion channel noise.} (A) and (B) Simulation results
from SHH model. lines are for guiding the eyes. S is the membrane
area. (C) and (D) Analytical solutions of Eq.\ref{single_pc} and
Eq.\ref{single_pr} from bistable model.} \label{fig1}
\end{figure}

The classic HH neuron model is a deterministic device. Action
potentials are caused only when it receives superthrshold stimulus.
In reality, ion channel noise could bring substantial effect on the
action potential initiation process. It makes the neuron fires in
response to subthreshold stimuli, or fail to fire in response to
supra-threshold stimuli\cite{White01071998}. To investigate the
effect of this so call threshold fluctuation on the signal detection
ability of neurons, we applied short (duration 1\emph{ms}) current
pulse inputs in the stochastic HH model, in which the membrane area
$S$ controls the number of ion channels, thus the intensity of ion
channel noise. If an action potential is generated immediately(less
than 8\emph{ms} after current injection), then we say the pulse
input is detected by the neuron, as demonstrated in Fig.~\ref{Demo}
. The pulse detection rate of the neuron is measured as the number
of pulse signal being detected over the total number of pulses
signal applied. In the absence of ion channel noise, i.e., the
membrane area is extremely large, the stochastic HH model is
equivalent to the deterministic HH model, which has a rigid
threshold. However, if the membrane area is limited, the stochastic
HH model shows ``softed" threshold in response to signals, as shown
in Fig.~\ref{fig1}(A). Becasue the ion channel noise could
facilitate threshold crossing of subthreshold signals, thus enable
the neuron to detect the subthreshold signals \cite{Adair14102003,
Stacey01092001}. The detection rate is increased as the membrane
area is decreased for subthreshold pulses. Meanwhile, ion channel
noise sabotages the neuron's detection ability for superthreshold
signals. Therefore, for superthreshold pulses, the detection rate
drops as the membrane area is decreased. Whereas for the threshold
signals, whose strength is just strong enough to make the neurons
fire in the absence of noise, its detection rate is $50\%$, no
matter how large the noise intensity. So for threshold pulses, the
detection rate is independent on the membrane area of the neuron.

 Beside threshold fluctuation, ion channel noise can also
trigger action potentials even in the absent of external input.
Fig.\ref{fig1} (B) demonstrated the firing rate of spontaneous APs
for neurons with different membrane area. It is seen that
spontaneous firing rate of action potentials decreases rapidly as
the membrane area increases, and when membrane area is larger that
$200\mu m^2$, the spontaneous firings become very rare.

\subsection*{Mimic the threshold fluctuation and spontaneous firings with bistable model}
A comprehensive theoretical and numerical analysis of the HH
equations with the inclusion of stochastic channel dynamics was done
Chow and White~\cite{Chow19963013}. It is shown that the system can
be approximated by a one-dimensional bistable Langevin equation,
which describes the dynamics of a particle in a double well
potential. Due to this fact, we use the following bistable model to
mimic the threshold fluctuation and spontaneous firing of a neuron:

\begin{equation}
\dot{x}=ax-x^{3}+n^{-1/2}\Gamma(t), \label{eq-1}
\end{equation}
where
\begin{equation}
<\Gamma(t)>=0; <\Gamma(t)\Gamma(t^{'})>=2\delta(t-t^{'}),
\end{equation}
$x$ is the position of the particle and analogous to the membrane
potential. The model has two minimal at $x_{1}=-\sqrt{a}$ and
$x_{2}=\sqrt{a}$, and a saddle point $x_{B}=0$. $n$ is taken here as
the number of ion channels of the neuron, and control the intensity
of ion channel noise~\cite{PhysRevE.70.011903}. We take the left
minimal of the double well as the resting state of the neuron, and
the saddle point as the threshold. The firing of the neuron is then
analogous to a particle surmounting the barrier and crossing to the
other minimum. We will take advantage of simplicity of this bistable
neuron model to have analytical description of its firing
probability for pulse signal under ion channel noise perturbation,
and spontaneous firing rate.

For the double well model at hand, suppose we have a transient pulse
signal $\Delta x$ that push the particle to a position $x$ near the
saddle point $x_{B}$, ie., $\Delta x=x-x_{B}$.  Following Lecar and
Nossal's approach of linearizing around the
threshold~\cite{lecar_theory_1971}, we can obtain the probability of
finding the particle in the right well after an enough long time(see
see \textbf{Models and Methods} for details), i.e., the probability
that a pulse input signal is detected by the neuron, which is
\begin{equation}
p_{c}=\frac{1}{2}[1+erf( \sqrt{an/2}\Delta x)],
 \label{single_pc}
\end{equation}
where $erf$ is so called error function. It is seen from
Fig.\ref{fig1}(B) that Eq.\ref{single_pc} well described the
threshold fluctuation due to ion channel noise observed from SHH
model.

 This spontaneous firing of action potential can be treated as a classic barrier escape problem with
additive thermal noise described by Eq.\ref{eq-1}. According to
Kramers's formula for escape
rate~\cite{Stochastic_Methods_Gardiner}, the spontaneous action
potential firing rate for this double well model is(see Appendix II
for detail)
\begin{equation}
p_{r}=\frac{\sqrt{2}a}{2\pi}e^{-a^{2}n/4}.
 \label{single_pr}
\end{equation}
It is seen from Fig.\ref{fig1}(D) that the spontaneous firing rate
of APs increases exponentially as the number of ion channels
decreasing.

\subsection*{The Energy Efficiency of Single Neuron}

Let us investigate the energy efficiency of a bistable neuron in a
task of detecting transient pulse train that compose $n_{all}$
pulses, with strength $\Delta x$, separated averagely by $\Delta t$.
For the bistable neuron at hand, let us suppose the energy cost for
neuron with only one ion channel to generate an AP is 1. Then the
neuron would generate $n_{all}p_{c}$ APs for correctly detecting of
$n_{all}p_{c}$ pulses, with the energy cost $n n_{all}p_{c}$. The
neuron also generates $\approx n_{all}\Delta t p_{r}$ spontaneous
APs due to noise perturbation, with energy cost $n n_{all}\Delta t
p_{r}$. Then the efficiency of energy use for the neuron,  can be
written as
\begin{equation}
Q=\frac{p_{c}}{n(p_{c}+\Delta t p_{r})}.
\end{equation}
or we can define the energy cost of correctly detecting one pulse
signal as
\begin{equation}
T=n(1+\Delta t \frac{p_{r}}{p_{c}}).
\end{equation}
Take $a$, $\Delta x$, and $\Delta t$ as parameters and $n$ as a
continuous variables, the condition for $T$ as a function of $n$ has
a local minimum, i.e., $Q$ has a local maxmum is
\begin{equation}
\frac{dT(n)}{dn}=0,\\ \frac{d^2 T(n)}{dn^2}>0 \label{condition}.
\end{equation}
The above problem can be solved easily with numerical method, the
results shows that in certain parameter region of $(a, \Delta x
\Delta t)$, there is a root $n^{*}$ for $\frac{dT(n)}{dn}=0$, at
which $\frac{d^2 T(n)}{dn^2}|_{n=n^{*}}>0$. To shows the results
explicitly, we directly calculate energy efficiency $Q$ as a
function of $n$ for parameters $a$, $\Delta x$ and $\Delta t$.

\begin{figure}[h]
  \centering
 \includegraphics[width=0.85\textwidth]{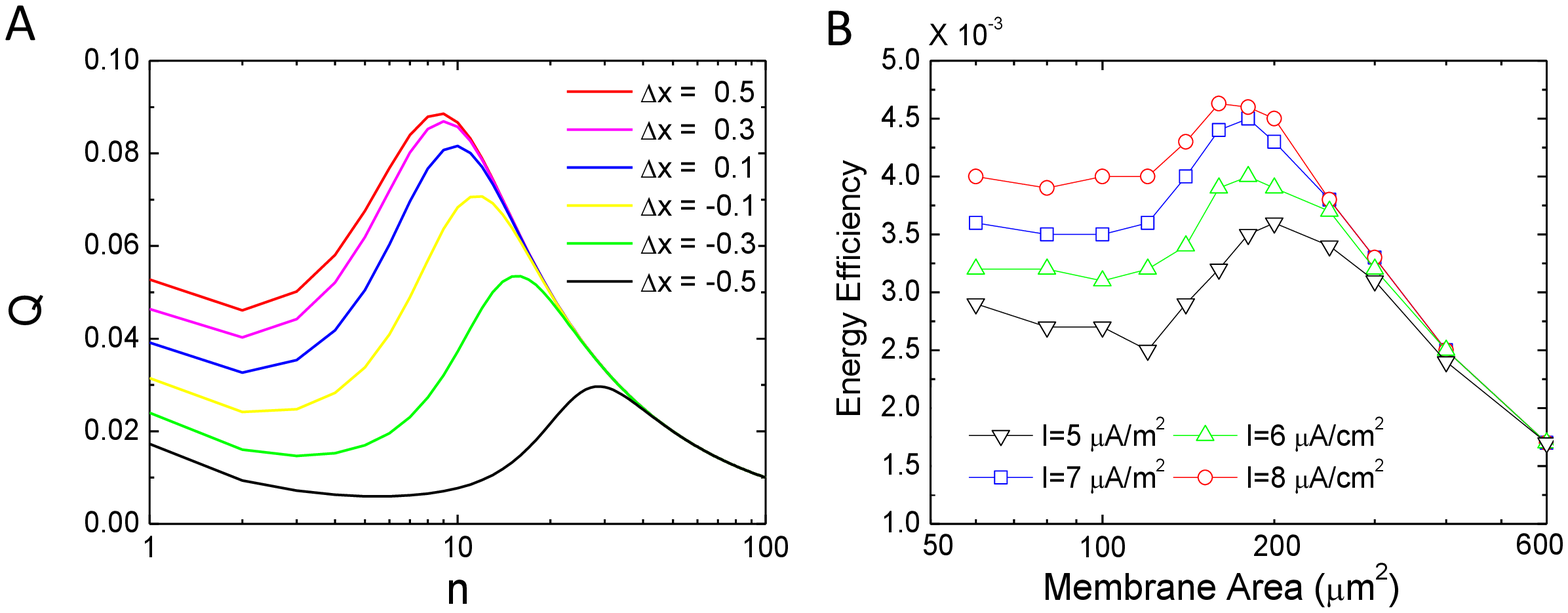}
\caption{ {\bf The Energy efficiency of a single neuron in the pulse
inputs detection task for different pulse strength.} (A) Energy
efficiency $Q$ as a function of number of ion channels $n$ of
bistable neuron model for different input pulse strength $\Delta x$.
$a=1$ and $\Delta t=100$. (B) Energy efficiency as a function of
membrane area of stochastic Hodgkin-Huxley neuron model for
different input pulse strength $I$. The inter-pulse interval is
$100ms$.} \label{fig2}
\end{figure}

 As shown in Fig.~\ref{fig2}(A), as $n$ increases, the energy efficiency increases to reach a maximum and then drops.
This resonance of energy efficiency with the number of ion channels,
i.e., noise intensity, implies there is an optimal number of ion
channels for the neuron that could take the pulse signal detection
task most efficiently in the energy consumption.

For the stochastic HH model, a pulse train that compose
$n_{all}=2000$ pulses, with strength $I$ and duration $1\emph{ms}$,
separated averagely by $\Delta t=100 ms$, is applied for detectoin.
The energy efficiency to detect this pulse train is defined as the
total number of pulse detected by the neuron over the total energy
costed in this detection task(see \textbf{Models and Methods} for
details). The results confirms that the energy efficiency do have a
maximum with proper input pulse interval $\Delta t$, as is seen from
Fig.~\ref{fig2}(B). It is seen that the energy efficiency exhibites
local maximum with membrane area around $180 \mu m^{2}$. Above this
optimal membrane are, the energy efficiency drops as the membrane
area increases. Below this optimal membrane area, the energy
efficiency drops to a minimum then increase as the membrane area
decreases. We argue this resonance of energy efficiency is actually
a balance between energy cost and pulse detection capacity. Because
large membrane area implies more ion channels are involved, thus
more ions need to be restored by the ATP pump after generation of
action potentials, to improve energy efficiency, neurons prefer
small membrane area. On the other hand, neurons with small membrane
area will generate more spontaneous APs in a certain period, which
contribute nothing to information transfer but cost energy.
Therefore, a minimum in energy cost, thus a maximin in energy
efficiency for a optimal membrane area is expected.

\subsubsection*{The maximal of energy efficiency depends on properties of pulse input}

The optimal number of ion channels for maximal energy efficiency is
dependent on the properties of the pulse train the neuron is
detecting.  As is seen from Fig.\ref{fig2}, neurons are tends to
detect strong pulse more reliably. Therefore, increasing the pulse
input strength, the energy efficiency increases, meanwhile the
optimal $n$ will decrease.

\begin{figure}[h]
   \centering
\includegraphics[width=0.85\textwidth]{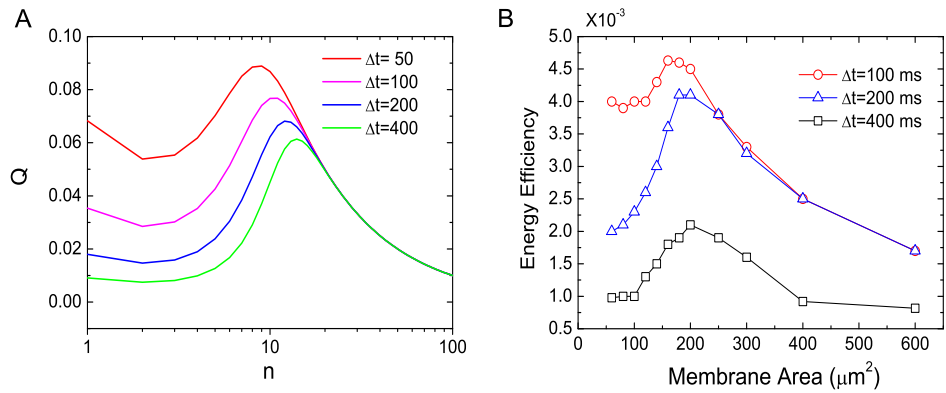}
\caption{ {\bf The Energy efficiency of a single neuron in the pulse
inputs detection task for different inter-pulse interval.} (A)
Energy efficiency $Q$ as a function of number of ion channels $n$ of
bistable neuron model for different inter-pulse interval $\Delta t$.
$a=1$ and $\Delta x=0.1$. (B) Energy efficiency as a function of
membrane area of stochastic Hodgkin-Huxley neuron model for
different input pulse strength $I$. The input pulse strength $I=8\mu
A/cm^{2}$. }
 \label{fig3}
\end{figure}

 The maximal energy efficiency is also dependent on the inter-pulse interval $\Delta t$
(Fig.~\ref{fig3}). Because in a large inter-pulse interval, the
neuron tends to fire more spontaneous APs, which contributing
nothing to signal detection but cost energy. Thus decreasing $\Delta
t$ decreases increases the energy efficiency, meanwhile the optimal
$n$ decreases. However, if $\Delta t$ is too small, the local
maximum would disappear and energy efficiency declines monotonously
with the number of ion channels.

\subsection*{The Energy Efficiency of Neuronal Population}
\begin{figure}[h]
\centering
\includegraphics[width=0.7\textwidth]{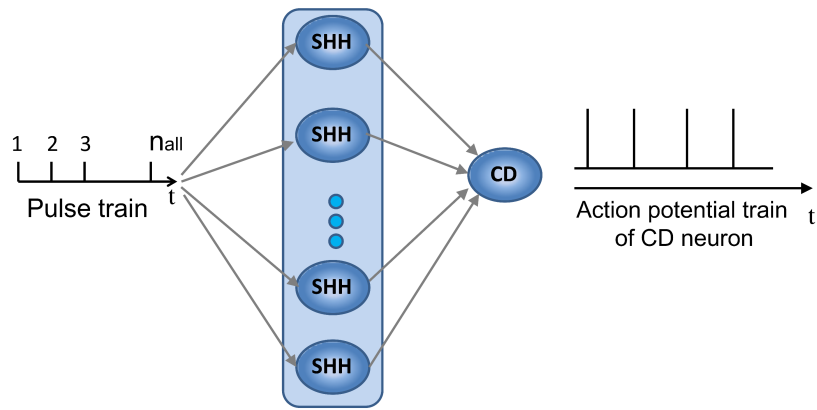}
\caption{ {\bf Pulse signal detection scenario for neural systems
with neuron population.} The front layer is composed with stochastic
neuron with ion channel noise(SHH), each of which receives the same
pulse train input. CD is coincidence detector neuron with threshold
$\theta$, it fires an action potential when $\theta$ or more than
$\theta$ front layer SHH neurons fire at the same time.}
\label{fig5}
\end{figure}

The neural system often transmit signals with synchronous spikes,
which has been shown that could prevent signals being damaged by
noise. The signal information can be reliably read out with the
coincidence-detector neuron. Here we use a simple scenario to
describe this process and investigate its energy efficiency. As is
shown in Fig. \ref{fig5}, a population of neurons receiving the same
pulse train input and send their output to a CD neuron. The CD
neuron takes outputs from neuron population as the input, and fires
only when multiple inputs arrive simultaneously in a short time
window $T_{w}$.

Suppose the CD neuron is excited by $\theta$ or more than $\theta$
inputs from $N$ neurons whose firing probability in response to
pulse stimulation is $p_{c}$, then the detection ability of this
neuron population, i.e., the firing probability of CD neuron, is
written as
\begin{equation}
P^{~\theta}(\theta, N, p_{c})= \sum _{\alpha =\theta}^N
\binom{N}{\alpha}p_{c}^{~\alpha}(1-p_{c})^{N-\alpha}=1-F( \theta,N,
p_{c})
 \label{p-theta}
\end{equation}
where $p^{~\alpha}(1-p)^{N-\alpha}$ is the probability that only
$\alpha$ neurons fire at the same time, and
$\binom{N}{\alpha}=\frac{N!}{\alpha!(N-\alpha)!}$ is the number of
ways of picking $\alpha$ neurons from population $N$.  So $\sum
_{\alpha =\theta}^N \frac{N!}{\alpha!(N-\alpha)!}$ is the total
number of ways of selecting $\theta$ or more than $\theta$ neuron
out of population $N$. $F(x , p, n)=\sum _{i
=0}^x\binom{n}{i}p^{i}(1-p)^{(n-i)}$ is the binomial cumulative
probability function.

We must note that in the the analysis of detection rate of single
neuron, we assumed the response time of single neuron to pulse input
could be infinitely large. However, in fact the most response
happens immediately after pulse input applied. Therefore, we could
expect a peak in the post-stimulus time histogram (PSTH) immediately
after stimulus, riding on the baseline of spontaneous firing(for an
example, see Fig. 2 (a) in Ref.~\cite{PhysRevE.78.051909}). So if
the CD time window $T_{w}$ is the same as the width of the peak in
PSTH, $p_{c}$ in Eq.~\ref{p-theta} could be approximated with
Eq.~\ref{single_pc}. In the following paragraph, we take this
approximation for the analytical solution of the CD neuron. And in
the simulation of stochastic HH model, we take CD time $T_{w}=8 ms$,
which is chosen to include most firings in response to the pulse
input, but not too long to include too much spontaneous
firings~\cite{10.1371/journal.pone.0056822}.

\subsubsection*{Detection ability enhanced with neuron population}
\begin{figure}[h]
\centering
\includegraphics[width=0.9\textwidth]{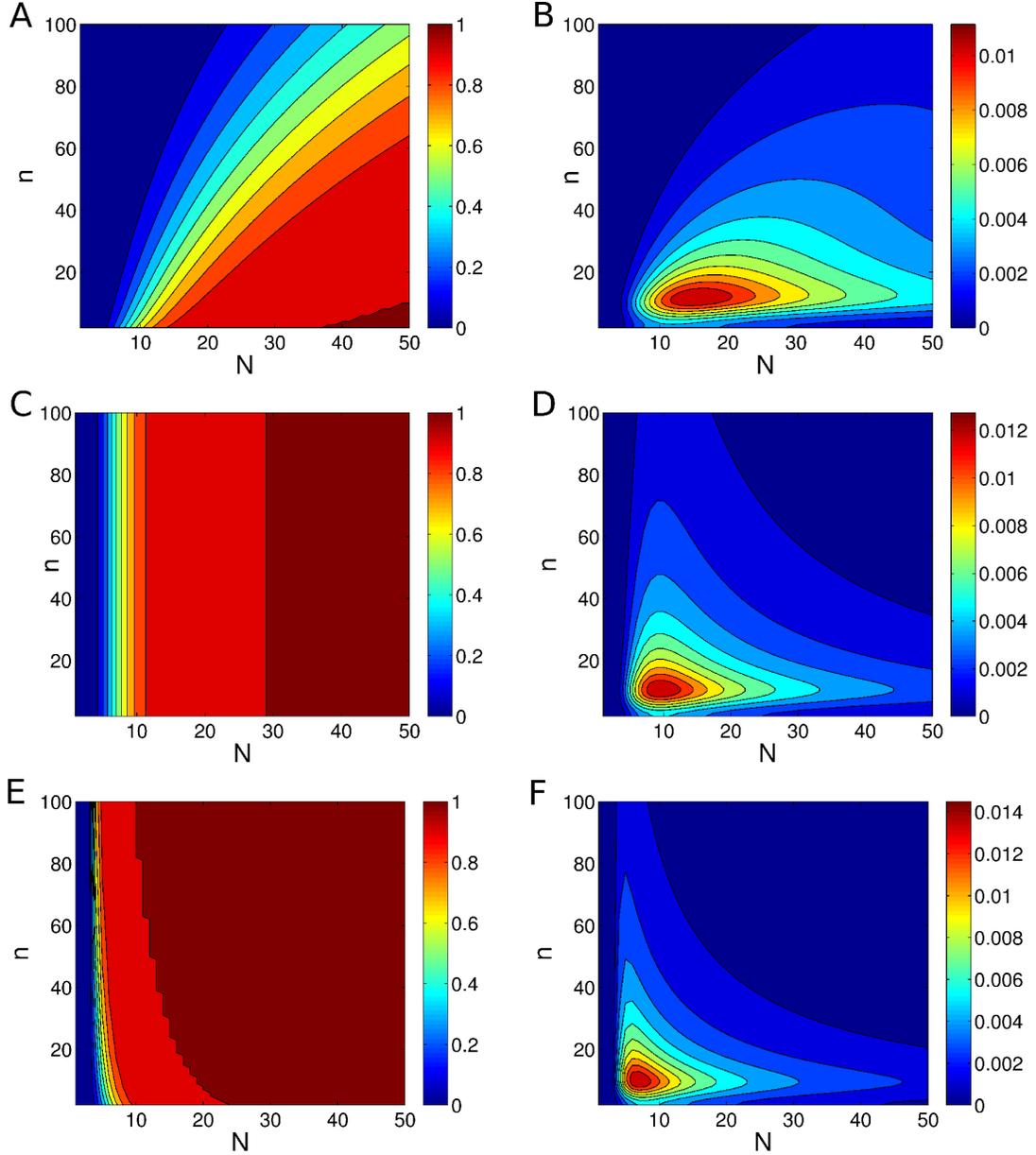}
\caption{{\bf Signal detection rate and energy efficiency for the
bistable neuron population for different input pulse strength.} Left
column corresponds to the signal detection rate and right column
corresponds to the energy efficiency. The top row corresponds to the
subthreshold pulse input:$\Delta x=-0.1$; The middle row corresponds
to threshold pulse input:$\Delta x=0$; The bottom row corresponds to
supthreshold pulse input:$\Delta x=0.1$. In all calculation,
parameters are $a=1$, inter-pulse interval $\Delta t=100$, the CD
threshold is $\theta =3$.} \label{fig6}
\end{figure}

Can detection ability be improved with neuron population? Combining
Eq.~\ref{single_pc} and Eq.~\ref{p-theta}, we calculated the
detection rate of CD neuron $P~^{\theta}$ for different size
configuration (n, N) of the upstream neuron population, i.e., the
number of ion channels in each neuron and the number of neurons in
the population. It is seen from Fig.~\ref{fig6} that for different
input pulse strength, there is a critical line in (n,N) plane. The
reliable detection ($P~^{\theta}\approx 1$) is realized in the area
on the right of critical lines. As the input pulse strength
increases, this critical line moves to the left in the (n,N) plane,
which implies the reliable detection could be realized with less
neurons in the population. For subthreshold or superthreshold pulse
inputs, this critical lines is depend on both the number of neurons
and the number of ion channels in neurons. For subthreshold pulse
input, the more ion channels in neurons, the lower the detection
rate of single neuron, thus the more neurons are needed to transmit
the input information to CD neurons to make correct
detection(Fig.~\ref{fig6}(A)). For superthreshold input, the more
ion channels in neurons, the higher the detection rate of single
neuron, thus the less neurons are needed to reach reliable
detection(Fig.~\ref{fig6}(E)). However, for threshold pulse input,
the critical line is independent of $n$(Fig.~\ref{fig6}(C)). Because
as discussed above, the detection rate of each single neuron is 0.5
for threshold pulse input, independent of the number of ion channels
in neurons.

The enhancement of detection rate of CD neuron in the circuit is a
consequence of higher firing probability of neurons in response to
pulse inputs in a time window $T_{w}$ than the spontaneous
firings~\cite{PhysRevE.78.051909}. The CD neuron is more sensible to
the nested high rate outputs of upstream neurons in response to
pulse stimuli than the randomly outputs of upstream neurons'
spontaneous firing. For upstream neurons with detection rate
$p_{c}$, if the number of neurons is large enough, the nested
firings due to pulse inputs are able to make CD neuron cross its
threshold $\theta$ and read out the information. Meanwhile, the
random spontaneous firings from upstream neurons are filtered by CD
threshold.

\begin{figure}[h]
\centering
 \includegraphics[width=0.7\textwidth]{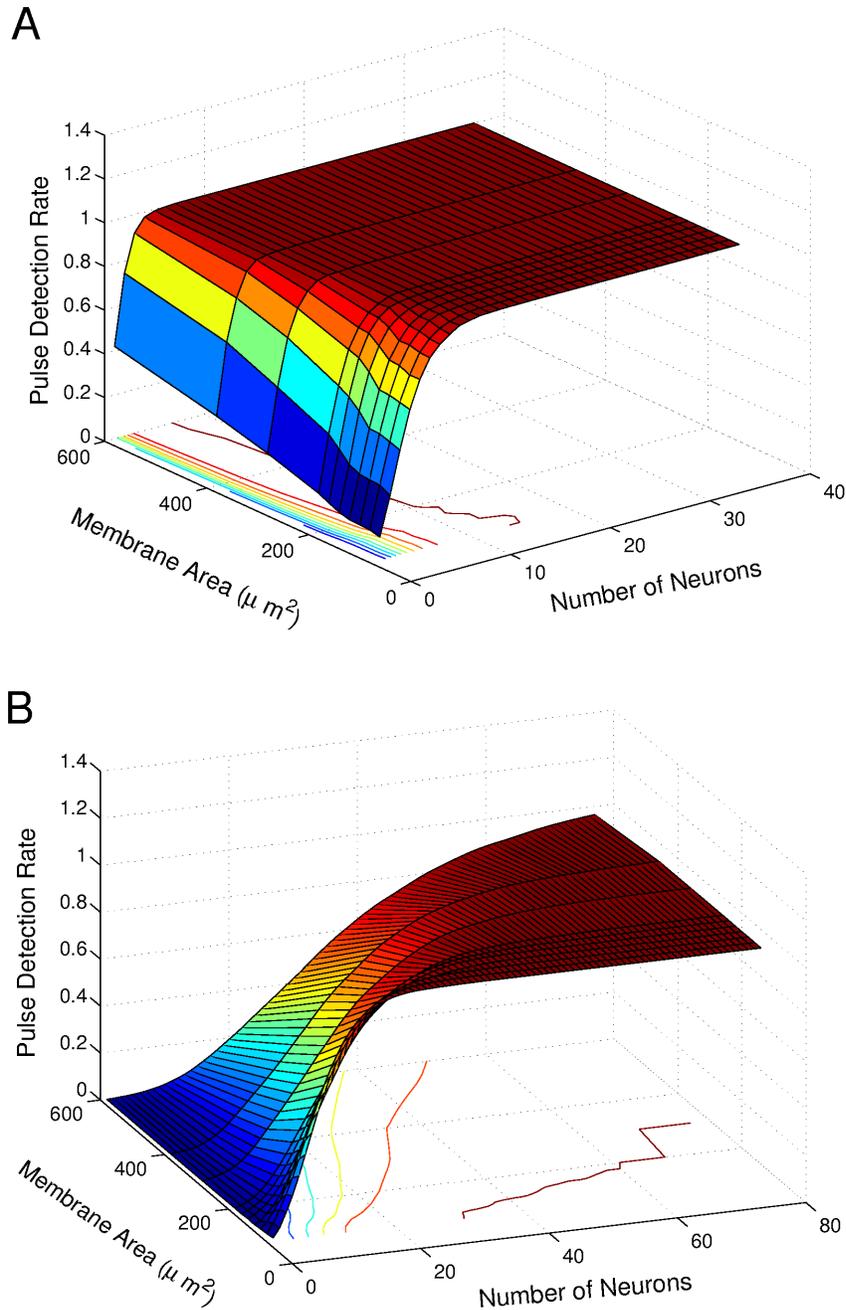}
\caption{ {\bf The pulse detection rate as a function of membrane
area in SHH model and the number of SHH neurons in neuron
population. } The input pulse strength is $8 \mu A /cm^{2}$ in (A)
and $5 \mu A /cm^{2}$ in (B). The inter-pulse interval is $100ms$.
The threshold of CD neuron $\theta=4$ and the time window for
coincidence detection is $8ms$.   } \label{pop_det}
\end{figure}

 For the simulation of stochastic HH model population, the same pulse train
composed by 2000 pulses (duration $1ms$) and separated averagely by
$100ms$ is applied to each neuron in the population. The output APs
of those neurons are converted to a point process in time, and
scanned by a sliding window of width $T_{w}=8ms$. If $\theta$ or
more than $\theta$ events are found in the sliding window, the CD
neuron is marked with a ``firing'' at the time the last event in the
window happens. Then the window starts to slid again after the last
event. The pulses are detected if there are ``firing''s in CD neuron
that happen less than $8ms$ after pulses applied. The detection rate
of the CD neuron is calculated as the number of pulses being
detected by the CD neuron over total number of pulses applied.
Simulation results in Fig.~\ref{pop_det} shows same behavior of
pulse detection rate depending on the size of the SHH neuron
population and the size of each SHH neuron.

\subsubsection*{Optimal size of the neuron population for energy efficiency }
 Next, we investigated the energy efficiency when input information is transmitted with
a population of neurons and read out by CD neuron.
 To detect the same pulse train described above,
this neuron population will
 response with $N n_{all}p_{c}$ APs
 in the population and $n_{all}P^{\theta}$ APs of CD neuron.
Meanwhile, there will be about $Nn_{all}p_{r}\Delta t$ spontaneous
APs in the population. Thus the neuron population detect
$n_{all}P^{\theta}$ pulses with energy cost of $n(Nn_{all}p_{c}+
Nn_{all}p_{r}\Delta t)$. Then the energy efficiency of this neuron
population is
\begin{equation}
Q=\frac{P^{\theta}(\theta, N, p_{c})}{Nn(p_{c}+ p_{r}\Delta t)}.
 \label{q-theta}
\end{equation}

Combining Eq.~\ref{single_pc}, Eq.~\ref{p-theta} and
Eq.~\ref{q-theta}, we calculated the energy efficiency for the
neuron population with different size configuration (n,N). We found
that the energy efficiency of the neuron population could be
maximized by the number of neurons. Therefore, there is an optimal
combination of the number of neurons and the number of ion channels
in neurons, with which the energy efficiency of the neuron
population is at its maximum (Fig.~\ref{fig6}, (B),(D),(F)). It
implies the energy efficiency could be optimized not only by the
number of the ion channels, but also the number of the neurons. We
argue that the optimal number of neurons $N^*$ is a compromise
between reliability of signal detection and the energy cost. As
discussed before, with the increasing of the number of neurons, the
detection rate of CD neuron increase, which gives an increasing in
energy efficiency. However, more neurons will generate more APs due
to pulse inputs, or spontaneous firings. As the number of neurons
increases, the energy cost increases too, which will decrease the
energy efficiency. When $N$ is large enough to cross the critical
lines for reliable detection, the detection rate becomes $1$ and no
longer depends on $N$, but the energy cost continue to increase
because redundance of spikes induced by input pulses, and the
spontaneous firing spikes.

\begin{figure}[h]
\centering
\includegraphics[width=0.7\textwidth]{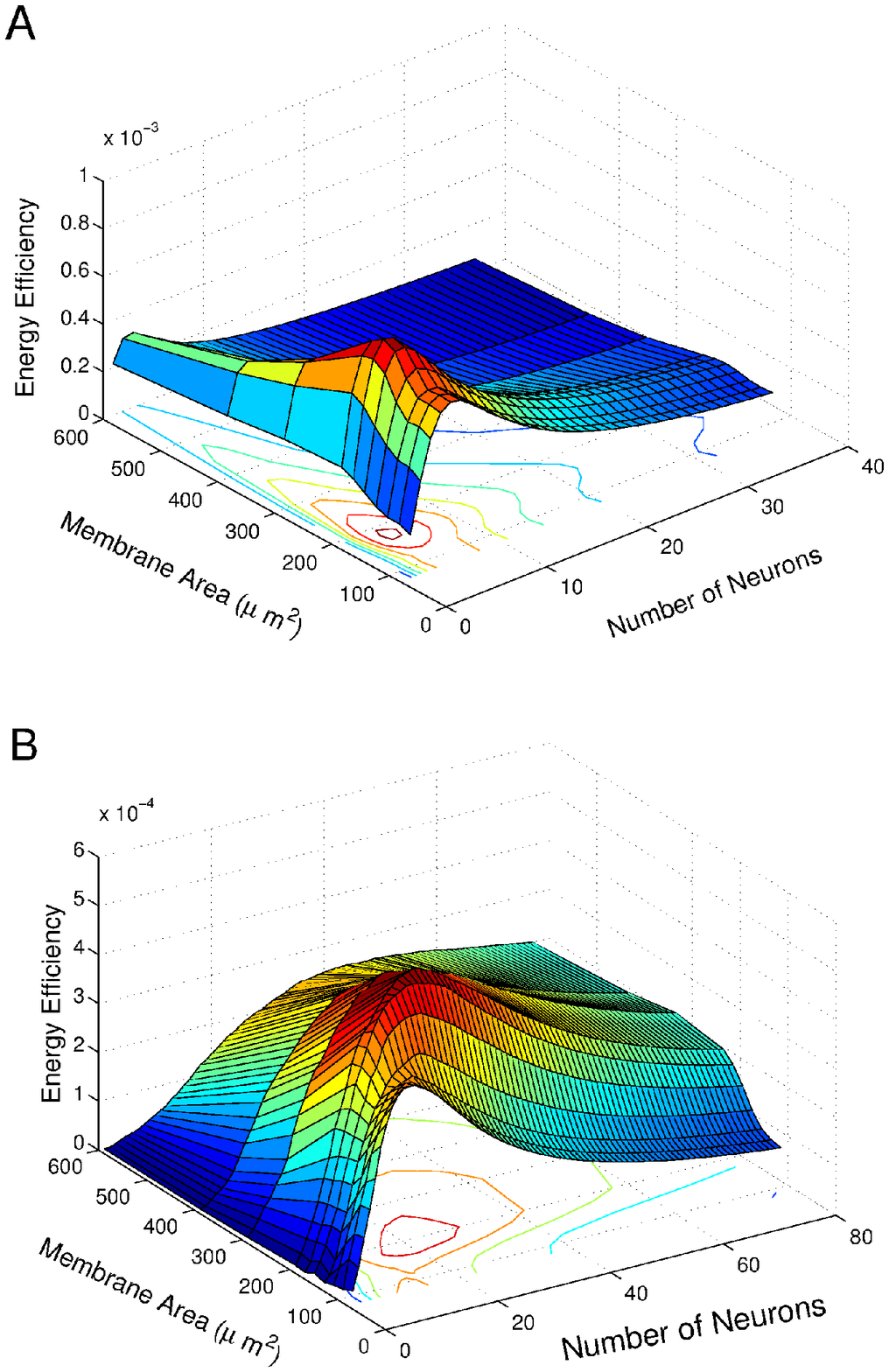}
\caption{ {\bf The energy efficiency as a function of membrane area
in SHH model and the number of SHH neurons in neuron population. }
The input pulse strength is $8 \mu A /cm^{2}$ in (A) and $5 \mu A
/cm^{2}$ in (B). The inter-pulse interval is $100ms$.
 The threshold of CD neuron $\theta=4$ and the time window for
coincidence detection is $8ms$.   } \label{fig7}
\end{figure}

  In the pulse detection process, the energy cost of the neuron population is calculated as the total
number of APs generated in the SHH neuron population, multiplied
with the membrane area of each neuron, assuming each AP generated in
unit membrane area cost one unit energy. Then the energy efficiency
is calculated as the detection rate of CD neuron over the energy
cost of neuron population. The simulation results shows that the
energy efficiency of the stochastic HH neuron population is
maximized with optimal combination of the number of ion channels
  and the number of neurons, as demonstrated in Fig.~\ref{fig7}.

\subsubsection*{Maximal energy efficiency depends on pulse strength and CD threshold}
\begin{figure}[h]
\centering
\includegraphics[width=0.7\textwidth]{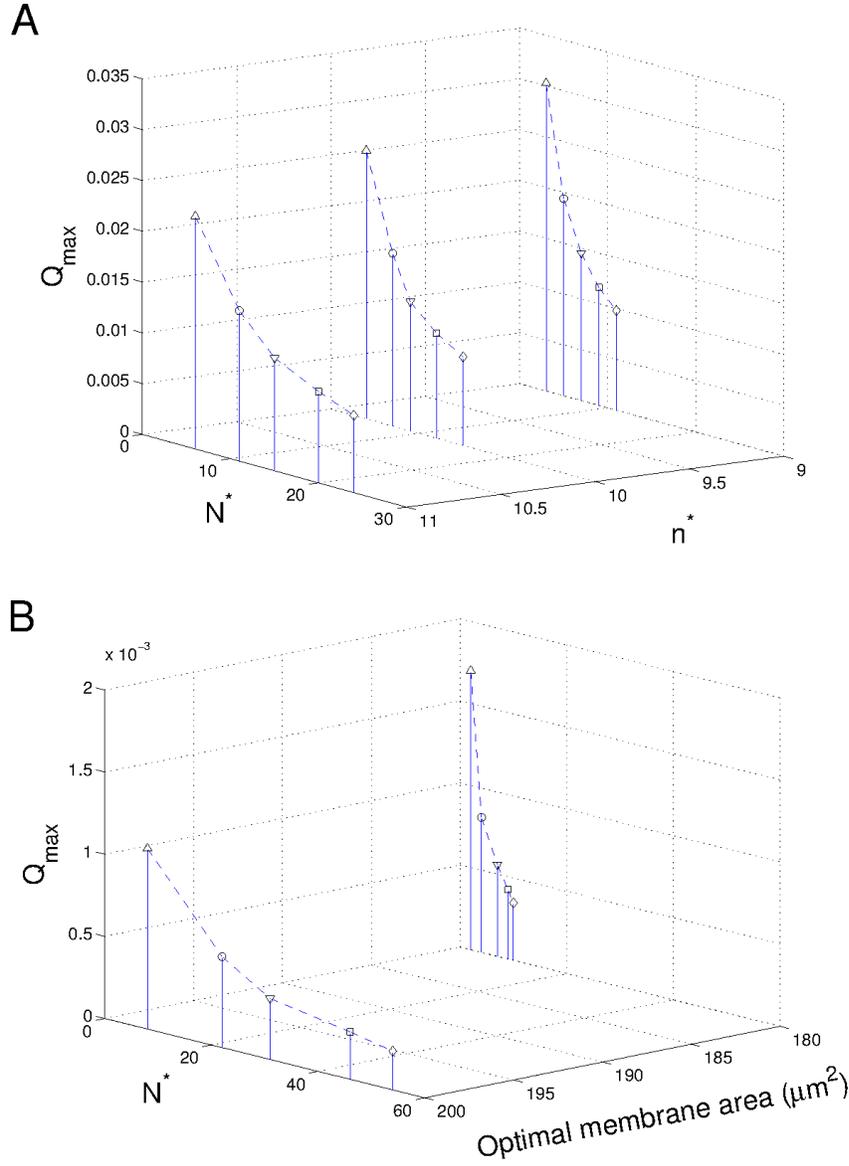}\\
\caption{ {\bf Dependence of the maximal energy efficiency and
corresponding optimal neuron population size on CD threshold and
input pulse strength. } (A) Analytical results for neuron population
with bistable model. From right to left, each dashed line
corresponds to $\Delta x =0.1, 0,-0.1$.  $\diamond$: $\theta
=5$;$\Box$: $\theta =4$;$\nabla $: $\theta =3$;$\circ$: $\theta
=2$;$\triangle$: $\theta =1$. Inter-pulse interval $\Delta t=100$.
(B) Simulation results for neuron network with SHH model. The right
line corresponds to $I=8$ and left $I=5$. $\diamond$: $\theta
=10$;$\Box$: $\theta =8$;$\nabla $: $\theta =6$;$\circ$: $\theta
=4$;$\triangle$: $\theta =2$. The inter-pulse interval is $100 ms$.
} \label{fig9}
\end{figure}

The optimal configuration of $(n,N)$ is dependent on the inter-pulse
interval(results not show), input pulse strength, and the
coincidence-detector threshold, as is seen from Fig.~\ref{fig9}. It
is seen that the maximal energy efficiency increases, as the input
pulse strength increases, meanwhile the optimal $(n,N)$
configuration moves towards the direction both $N$ and $n$
decreases.

We can also find in Fig.~\ref{fig9} that for the same input pulse
strength, as the coincidence detector threshold increases, the
optimal $(n,N)$ configuration moves towards the direction $N$
increases, independent of $n$.  Indeed, CD threshold has no effects
on the detection rate of upstream neurons, but large CD threshold
requires more synchronous spikes induced by the input pulses to make
correct detection. We must note that the role of CD neuron in our
study is to enhance the detection rate for pulse signals, but it has
another role as suppressing the transduction of spontaneous firings.
Because spontaneous firings happens randomly in time and has little
chance to be synchronized, it will be filtered by the CD
neuron~\cite{10.1371/journal.pone.0056822}. Therefore, although our
results point out that the lower CD threshold gives higher energy
efficiency, for the purpose of suppressing the spontaneous firings,
the CD threshold should be larger than 1.

\section*{Discussion}

Neural systems employ action potentials to carries input
information. The generation and propagation of action potentials are
supported by metabolic energy, which is a large burden for animals
and is believed to shape the neural system through evolution
pressure. Here we investigated the energy efficiency of neural
system performing a pulse signal detection task under the
perturbation of ion channel noise. We found the energy efficiency
exhibits peak values, corresponding to an optimal sizes of the
neural system, i.e., number of the ion channels in the neuron,
number of neurons in the network.

The optimal size of the neurons for maximal energy efficiency is
actually a balance between reliability of signal processing and
energy cost, as previously pointed out by Schreiber \emph{et.al}
with linear input-output neurons composed by sodium and potassium
channels~\cite{Schreiber02energy-efficientcoding}. In our study, we
showed this principle holds if neurons use action potentials to
transfer information. Furthermore, we also showed that this
principle also holds for neuron population, and the number of
neurons is optimized for maximal energy efficiency.

The intrinsic ion channel noise, which is inversely proportional to
the number of ion channels, play a double-edged sword role in the
signal process of neurons, as well studied in the context of
stochastic resonance\cite{McDonnell1000348, Wiesenfeld1995}. In this
paper we show that it is also important in the energy efficiency of
neurons in signal processing. Ion channel noise helps neurons to
detect subthreshold signals, thus improves the energy efficiency.
Meanwhile, it damages the reliability of neurons for superthreshold
signal detection, and generate spontaneous action potentials, thus
decreases the energy efficiency. Form the deduction of detection
rate and spontaneous firing rate for the bistable neuron model, we
see that the resonance of energy efficiency is a kind of stochastic
resonance. For example, the maximal energy efficiency exists if we
use other kind of noise, e.g., synaptic noise, which is caused by
many independent presynaptic current. So we expect, in general,
energy efficiency of single neuron could be maximized for certain
level of noise, and the number of neurons in the noisy environment
could be optimized for maximal energy efficiency.

In this study, we evaluated the energy efficiency in a simple coding
scenario of neural systems, in which the input pulse signals are
encoded digitally with the action potentials. we concluded that the
neural system size could be optimized for maximal energy efficiency.
In a general way, the information conveying ability of neural
systems could be measured by information theory formulated by
Shannon~\cite{SW49}. Then this rule could be rechecked in this
general framework by defining the energy efficiency as the ratio
between the amount of transmitted information and the metabolic
energy required~\cite{PhysRevE.83.031912}. The analysis in our work
provides a starting point on this direction and can guide further
experimental and theoretical work.

\section*{Models and Methods}
\subsection*{The Stochastic Hodgkin-Huxley model}

\begin{figure}[h]
\centering
\includegraphics[width=0.5\textwidth]{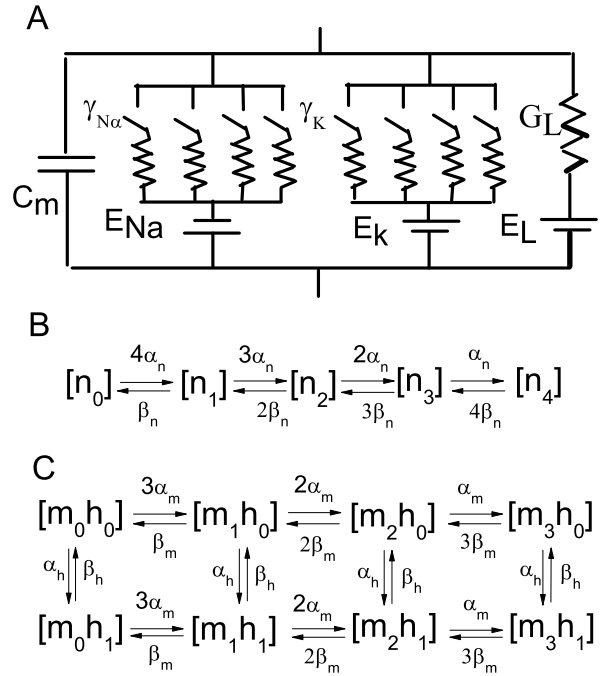}
\caption{ {\bf The scheme of stochastic Hodgkin-Huxley model.} (A)
Electrical equivalent circuit for the  Hodgkin-Huxley Model with
consideration of each ion channel conductance. (B) The Markov chain
model for potassium channels. Channels are opened when they are in
$[n_{4}]$ state and closed in other states. (C) The Markov chain
model for sodium channels. Channels are opened when they are in
$[m_{3}h_{1}]$ state and closed in other states.} \label{HHcircuit}
\end{figure}

According to classic HH model, the current flowing across the giant
axon membrane is represented by the sum of the capacitative current
and the ionic currents through the corresponding conductive
components, which are averaged deterministic terms assuming the
numbers of ion channels are large ~\cite{Hodgkin}. For limited
numbers of ion channels, the total ion conductances for $Na^{+}$ and
$K^{+}$ should be replaced with the single-channel conductances of
potassium and sodium channels, resulting in equivalent circuit
demonstrated in Fig.~\ref{HHcircuit}(A). The membrane dynamics of
the HH equations is then given by
\begin{eqnarray}
C_{m}\frac{dV}{dt} &=& -(G_{K}(V-V_{K}^{rev})+ \;G_{Na}(V-V_{Na}^{rev})\nonumber\\
 & &+\;G_{L}(V-V_{L}))+\,I,
\label{eq-HH}
\end{eqnarray}
where $V$ is the membrane potential and $I$ the input current.
$V_{K}^{rev}$, $V_{Na}^{rev}$, and $V_{L}$ are the reversal
potentials of the potassium, sodium and leakage currents,
respectively. $G_{K}$, $G_{Na}$, and $G_{L}$ are the corresponding
specific ion conductances, and $C_{m}$ is the specific membrane
capacitance. The conductances for potassium and sodium channel are
given by
\begin{equation}
G_{K}(V,~t)=\gamma_{K}[n_{4}]/S \;,\;
       G_{Na}(V,~t)=\gamma_{Na}[m_{3}h_{1}]/S\;,
 \label{eq-5}
\end{equation}
where $\gamma_{K}$ and $\gamma_{Na}$ are the single-channel
conductances of potassium and sodium channels. $[n_{4}]$ refers to
the number of open potassium ion channels and $[m_{3}h_{1}]$ refers
to the number of open sodium ion channels. $S$ is the membrane area
of a neuron. The total number of sodium channels and potassium
channels are given by $\rho_{Na}S$ and $\rho_{K}S$, where
$\rho_{Na}$ and $\rho_{K}$ are the densities of sodium and potassium
channels. In this work, we use $S$ to control the number of ion
channels, thus the ion channel noise intensity.

  The gating dynamics of each ion channels are modeled with corresponding Markov process, as demonstrated
  in Fig.~\ref{HHcircuit}(B)) and Fig.~\ref{HHcircuit}(C). The
  $K^{+}$ channels can exist in five different states and switch between these
states according to the voltage dependence of the transition rates,
and the channels opens only when it is in $n_{4}$ state. Similarly,
the $Na^{+}$ channel has eight states, with only one open state
$m_{3}h_{1}$. In each time step, the number of open $K^{+}$ channels
and $Na^{+}$ channels are determined by stochastic simulation of
Markov chain model of those channels, and Eq.~\ref{eq-HH} is
integrated by Euler-forward method with time step $0.01ms$. The
parameters and rate functions used in the simulation of stochastic
HH model are listed in Table S1.

\subsection*{Measure the energy cost of stochastic HH neuron model}
Usually, the energy cost of a neuron can be calculated by
integrating the sodium current or potassium current over the time,
then convert into the number of sodium or potassium ions pumped in
and out of the cell. Then energy cost is obtained if the powers the
$Na^{+}/K^{+}$ pump is measured from experiment(for example, 50
kJ/mol in
heart~\cite{CrottySeptember_2006})~\cite{10.1371/journal.pcbi.1002456}.
However, this method is not efficient when the energy cost of a
neuron is estimated in a long time intervals. In fact, the shape of
the action potential is stereotyped, and the currents flow in and
out of the membrane are described in unit area in the model. So in
an action potential generation process, the energy cost in unit area
is a constant, independent of input signals. Therefore, in this
study, we counted the number of action potentials generated in the
whole pulse detection
 process, then multiplied with the membrane area of the neuron as the
 measurement of energy cost for this neuron. With this approach, we ignored the
 energy cost in the subthreshold fluctuation of membrane potential caused by ion channel
 noise. Because the amplitude of subthreshold fluctuation is very small comparing to the amplitude of action potentials~\cite{Steinmetz99subthresholdvoltage}, the energy cost of subthreshold fluctuation is trivial
 comparing to the energy cost of action potentials.

\subsection*{Bistable neuron model and its response function to pulse input}
For a neuron subjects to a pulse signal and noise, its dynamics is
analogous to a particle in a double well potential with pulse force
and noise perturbation, thus could be described with the following
equation:
\begin{equation}
\dot{x}=-U^{'}(x)+\Gamma(t),
 \label{eq-A1}
\end{equation}
where
\begin{equation}
<\Gamma(t)>=0; <\Gamma(t)\Gamma(t^{'})>=2D\delta(t-t^{'}).
\end{equation}
$U=-\frac{a}{2}x^2+\frac{x^4}{4}$ is a double well potential, which
has two minimal at $x_{1}=-\sqrt{a}$ and $x_{2}=\sqrt{a}$ and a
saddle point $x_{B}=0$. $D$ is noise intensity. Let's assume that a
short duration force moves the particle parallel to the x-axis into
the region of the saddle point. After the force is removed, the
particle drifts up to the region of the saddle point. Near the
saddle point, the phase trajectories are repelled, causing the
particle to accelerate away from the saddle- point region towards
one of the minimal. Thus we can calculate the probability that,
after a long time, the particle is found in the $x>x_{B}$ domain.

Next we expand Eq.~\ref{eq-A1} in the neighborhood of the threshold
singular point $x_{B}$. Letting
\begin{equation}
\epsilon = x-x_{B}
\end{equation}
and remember $x_B=0$, we obtain the equation:
\begin{equation}
\dot{\epsilon} =a\epsilon +\Gamma(t) \label{a2}
\end{equation}
the solution of Eq.~\ref{a2} has the form
\begin{equation}
\epsilon (t)=\epsilon (0)e^{at}+\int^{t}_{0}e^{a(t-s)}\Gamma(s)ds
\end{equation}
This integral can be rewritten as:
\begin{equation}
X(t)=\epsilon
(t)-\epsilon(0)e^{at}=\int^{t}_{0}e^{a(t-s)}\Gamma(s)ds
 \label{a4}
\end{equation}
Since $\Gamma(t)$ is a gaussian random variable. It follows that the
time integral $X(t)$ also obey a gaussian distribution. Then we have
\begin{equation}
P(X,t)=(2\pi<X^{2}>)^{-\frac{1}{2}}exp(\frac{-X^{2}}{2<X^{2}>})
\end{equation}
where the angular bracket denote an expectation value. The
expectation value of of $X^{2}$ can be expressed in terms of the
joint expectation of the variable $\Gamma$ taken with itself at a
different time. From Eq.~\ref{a4} one finds
\begin{equation}
<X^{2}>=\frac{D}{a}(e^{2at}-1).
\end{equation}
The probability of the neuron firing after a pulse input is equal to
the probability that $\epsilon(t)>0 $ when $t\longrightarrow
\infty$, given an initial displacement $\epsilon (0)$, Then
\begin{equation}
P[\epsilon (t)>0|\epsilon(0)]=\lim_{t\rightarrow
\infty}\int^{\infty}_{-\epsilon(0)e^{at}}P(X,t)dX
\end{equation}
or
\begin{equation}
P[\epsilon
(t)>0|\epsilon(0)]=\frac{1}{2}[1+erf(\frac{\epsilon(0)}{\sqrt{2D/a}})]
\end{equation}
where \emph{erf} is the so-called error function, defined as
\begin{equation}
erf(x)=\frac{2}{\sqrt{\pi}}\int^{x}_{0}e^{-t^{2}}dt.
\end{equation}

\subsection*{The Spontaneous firing rate of bistable neuron model}
The spontaneous firing rate of a bistable neuron is considered as
the escaping rate of a paretical form right well to the left well.
According to Kramers's formula, the escaping rate is
\begin{equation}
K=\frac{1}{2\pi}\sqrt{U''(x_{s})|U''(x_{u})|}exp(-\frac{\Delta
U}{D}),
\end{equation}
where $\Delta U=U(x_{u})-U(x_{s})$. $x_{u}$ and $x_{s}$ are the
position of stable state and the threshold the particle will cross,
respectively. $x_{u}=0$ and $x_{s}=-\sqrt{a}$ in our case.
Therefore, the escaping rate of a particle in the double well
potential is
\begin{equation}
K'=\frac{\sqrt{2}a}{2\pi}exp(-\frac{-a^2}{4D}).
\end{equation}

\section*{Supporting Information}
\textbf{Table S1 } Parameters and Rate Functions in the Stochastic
Hodgkin-Huxley Model.

\section*{Acknowledgments}
This work was supported by the National Natural Science Foundation
of China (Grants No. 11105062), the Fundamental Research Funds for
the Central Universities (Grant No. lzujbky-2011-57), the
Fundamental Research Funds for the Central Universities of Northwest
University for Nationalities(No zyz2011051). We thank Prof. Yuguo Yu
for his valuable comments on the manuscript.

\section*{Author contributions}
Conceived and designed the experiments: LCY. Performed the
experiments: LCY LWL. Analyzed the data: LCY LWL. Wrote the paper:
LCY.

\bibliography{ref}

\newpage

\section*{Figure Legends}

\end{CJK}

\end{document}